\pdfoutput=1
\documentclass[twocolumn,english,aps,superscriptaddress,prb]{revtex4-1}

\usepackage{babel}
\usepackage{amsmath}
\usepackage{amssymb}
\usepackage{graphicx}

\usepackage{placeins}
\usepackage{bm}% bold math

\begin{document}

\title{Critical properties of the quantum Ashkin-Teller chain with chiral perturbations}

\author{Bernhard E. L\"uscher}
\affiliation{Institute of Physics, University of Z\"urich, R\"amistrasse 71, 8006 Z\"urich, Switzerland}
 \author{Fr\'ed\'eric Mila}
 \affiliation{Institute of Physics, Ecole Polytechnique F\'ed\'erale de Lausanne (EPFL), CH-1015 Lausanne, Switzerland}
\author{Natalia Chepiga}
\affiliation{Kavli Institute of Nanoscience, Delft University of Technology, Lorentzweg 1, 2628 CJ Delft, The Netherlands}

\date{\today}
\begin{abstract} 
We investigate the nature of the phase transitions in the quantum Ashkin-Teller chain in the presence of chiral perturbations. We locate the Lifshitz line separating a region of direct chiral transitions from the region where the transition is through an intermediate floating phase. Furthermore, we identify a small region in the vicinity of the four-state Potts point where chiral perturbations are irrelevant and where the transition remains conformal. Implications to Rydberg atoms experiments are briefly discussed. 
%We examine the phase diagram of the Quantum Ashkin-Teller model with a chiral perturbation using a DMRG algorithm. As is the case in its classical equivalent, for a non-vanishing chiral perturbation the transition out of a commensurately ordered phase with wave vector $q=2\pi$ into a disordered phase is observed to happen via a floating phase as well as via a direct chiral transition. In addition, we report the discovery of a small region in the phase diagram with non-vanishing chiral perturbation where the transition remains conformal and occurs in the Ashkin-Teller univresality class. 
\end{abstract}

\maketitle

%%%%%%%%%%%%%%%%%%%%%%%%%%%%%%%%%%%% INTRODUCTION %%%%%%%%%%%%%%%%%%%%%%%%%%%%%%%%%%%%

\section{Introduction}

The theory of quantum phase transitions in low-dimensional systems plays a central role in modern condensed matter physics\cite{giamarchi,tsvelik}. One of the most challenging and long-standing problems is the commensurate-incommensurate (C-IC) melting of a period-$p$ phase originally formulated in the context of absorbed monolayers\cite{HuseFisher,Selke1982,schulz1983phase,haldane_bak,HuseFisher1984,SelkeExperiment}. 
Huse and Fisher noticed that domain walls between different ordered domains\cite{HuseFisher,HuseFisher1984}, for instance $A|B$ and $B|A$, might have different free energy contributions, introducing chiral perturbations into the problem. For the period-2 phase these chiral perturbations are irrelevant and the transition is in the Ising universality class. For period-$p$ phases with $p\geq 5$ the transition is always through an intermediate critical phase separated from the ordered phase by a Pokrovsky-Talapov\cite{Pokrovsky_Talapov} and from the disordered phase by a Kosterlitz-Thouless\cite{Kosterlitz_Thouless} transition.

The most interesting cases, however, are transitions out of the period-3 and period-4 phases. In the absence of chiral perturbations the transition out of the $p=3$ phase is in the 3-state Potts universality class that, as the Ising transition, can be described by a corresponding minimal model of the conformal field theory\cite{difrancesco}. 
For strong chiral perturbations, the transition is a two-step process via an intermediate critical phase, as for $p\geq 5$.
% via a Pokrovsky-Talapov (PT) and a Kosterlitz-Thouless (KT) transition respectively, with an intermediate floating phase in between the two. 
However, in the presence of weak chiral perturbations, the transition is believed to be a direct one in a new chiral universality class\cite{HuseFisher1984}. The melting of the $p=4$ phase, encapsulated by the physics of the Ashkin-Teller model, is even more complicated. Again, in the absence of chiral perturbation, the transition belongs to the Ashkin-Teller universality class that can be described by a conformal field theory with central charge $c=1$\cite{difrancesco}. However, in this case weak chiral perturbations may or may not give rise to a direct chiral transition\cite{Den_Nijs} depending on the properties of the Ashkin-Teller point itself. For instance, in the limit of the Ashkin-Teller model equivalent to two decoupled Ising chains, a chiral perturbation immediately opens up an intermediate incommensurate Luttinger liquid phase with a diverging correlation length, a phase known as a floating phase\cite{Ostlund,nelson1979dislocation}. In the opposite limit equivalent to the symmetric four-state Potts model, field theory arguments predict that weak chiral perturbations are irrelevant and that the transition remains direct and conformal\cite{schulz1983phase}.
 But what happens between these two limits? According to Huse and Fisher\cite{HuseFisher1984} there might be room for a chiral transition and numerical results in the context of Rydberg atoms have given the first evidence that this is indeed the case\cite{chepiga2021kibble}.  
This scenario is further supported by numerical simulations of the classical symmetric two-dimensional (2D) Ashkin-Teller model on a square lattice where the Lifshitz line - the boundary of chiral transition - was accurately located with a Corner Transfer Matrix Renormalizationgroup (CTMRG) algorithm \cite{nyckees2022commensurate}. 
% In quantum 1D chain chiral transition has been reported at the boundary of the period-4 phase in an effective blockade model of Rydberg atoms\cite{chepiga2021kibble}.
% <------ Probably this should be mentioned once it has been stated that the classical 1d model corresponds to the quantum 2d model in the strongly anisotropic limit?
Yet another feature of the melting out of a $p=4$ phase - the possibility of a conformal transition for a non-vanishing chiral perturbation - remains unexplored.  If the transition remains conformal even in the presence of chiral perturbations, the dynamical critical exponent $z$ must keep its value $z=1$, and there must be a quantitative correspondence between the classical 2D chiral Ashkin-Teller model on a square lattice and its quantum 1D version on a chain. 
% and is the main focus of the present paper.We provide numerical evidences of this scenario and define a scale of what can be considered as {\it weak} chiral perturbations.

Recent progress in Rydberg atoms experiments\cite{bernien2017probing,keesling2019quantum} brings this old problem into the main focus of theoretical research now in the context of quantum 1D chains. The phase diagram of a Rydberg atoms array is dominated by lobes of integer periodicities $p=2,3,4,5...$\cite{keesling2019quantum,rader2019floating} while the distance-dependent van der Waals interaction makes the disorder phase surrounding these ordered lobes incommensurate. This makes a 1D array of Rydberg atoms an ideal playground to probe quantum commensurate-incommensurate melting\cite{PhysRevB.98.205118,prl_chepiga,scipost_chepiga,dalmonte,sachdev_dual,PhysRevResearch.3.023049,chepiga2021kibble,PhysRevResearch.4.043102}. Numerical simulations of quantum phase transitions out of the period-4 lobe in Rydberg chains show that they are qualitatively very similar to the transition out of the period-3 phase\cite{chepiga2021kibble}: the conformal transition is realized at a single point and is surrounded by a direct chiral transition before the floating phase opens. The new developments in Rydberg atoms open the way to tune the conformal point within the Ashkin-Teller family and, in principle, can bring it to the point where the floating phase opens up immediately or to the point where the conformal transition extends to a finite interval.

\begin{figure}
\includegraphics[width=0.45\textwidth]{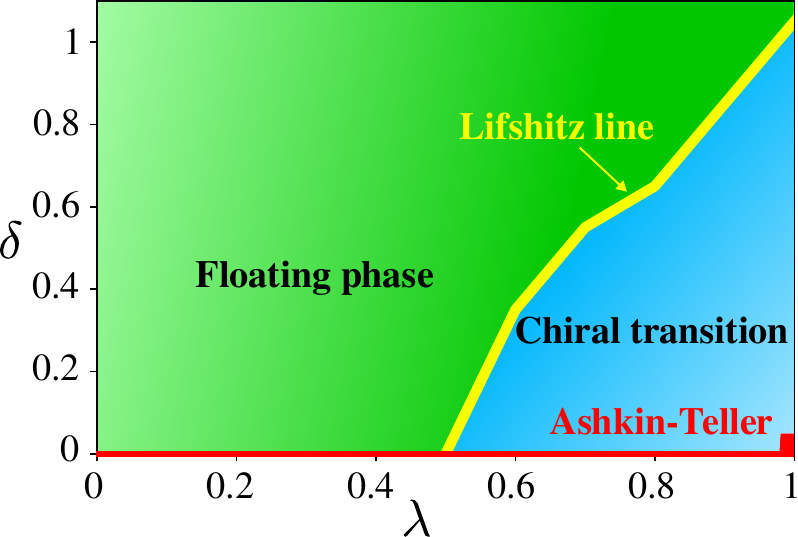}

\caption{Nature of the quantum phase transition between the ordered phase with a four-fold degenerate ground state and the disordered phase of the quantum Ashkin-Teller model defined by the Hamiltonian \ref{ChiralHamiltonian} as a function of $\lambda$ and $\delta$. Along $\delta=0$ the transition is in the Ashkin-Teller universality class. For $\lambda\lesssim 0.5$ we observe an intermediate floating phase (green) for any non-zero value of chiral coupling $\delta$. For $\lambda\gtrsim 0.5$ the transition is direct in the chiral universality class (light blue) below the Lifshitz line (red) and through the floating phase above it. For $\lambda\gtrsim 0.977$ we find that the transition remains direct and conformal in the Ashkin-Teller universality class (dark blue) for $\delta\lesssim 0.04$.}
\label{fig:Diagram1}
\end{figure}

In the present paper we provide numerical evidence that a conformal transition can be realized even in the presence of weak chiral perturbations and define a scale on which interactions can be treated as {\it weak} for the quantum Ashkin-Teller model. We also locate the Lifshitz line, i.e. the position in the phase diagram where the direct chiral transition is replaced by the floating phase. Our main results are summarized in the phase diagram presented in Fig.\ref{fig:Diagram1}. The rest of the paper is organized as follows. In Section \ref{sec:method} we define the Ashkin-Teller model and provide the details of the numerical algorithm. In Section \ref{sec:results} we present our results including the detailed overview  of the phase diagram and we provide numerical evidence in favour of a conformal transition in the presence of non-zero chiral perturbations. In Section \ref{sec:discussion} we summarized the results and put them into perspective.

\section{Model \& Methods}
\label{sec:method}
\subsubsection{The Quantum Ashkin-Teller Model}
In the quantum Ashkin-Teller model two Ising spins per site, $\hat{\sigma}_i$ and $\hat{\tau}_i$, are introduced and embedded into a Hamiltonian of the form
\begin{equation}\label{Hamiltonian}
    \begin{split}
        \hat{H}_0=-&\sum_{i}\left(\hat{\sigma}^{x}_i + \hat{\tau}^{x}_i+\lambda\hat{\sigma}^{x}_i\hat{\tau}^{x}_i \right)\\
        -\beta&\sum_i\left(\hat{\sigma}^{z}_i\hat{\sigma}^{z}_{i+1}+\hat{\tau}^{z}_i\hat{\tau}^{z}_{i+1}+\lambda\hat{\sigma}^{z}_i\hat{\sigma}^{z}_{i+1}\hat{\tau}^{z}_i\hat{\tau}^{z}_{i+1}\right).
    \end{split}
\end{equation}
This model displays Kramers-Wannier self-duality at $\beta=1$.  This point corresponds to a quantum phase transition between an ordered phase at large $\beta$ and a disordered phase at small $\beta$\cite{kohmoto}. It is in the Ashkin-Teller universality class with exponents that vary continuously with $\lambda$. 
% Introducing the operators $\hat{P}_1=\Pi_i\hat{\sigma}_i^{x}$, $\hat{P}_2=\Pi_i\hat{\tau}_i^{x}$, and $\hat{P}_3=\Pi_i\hat{\tau}_i^{x}\hat{\sigma}_i^{x}$ it is easily verified that $\left[\hat{H}_0,\hat{P}_1 \right]=\left[\hat{H}_0,\hat{P}_2 \right]=\left[\hat{H}_0,\hat{P}_3 \right]=0$, indicating the 4 fold degeneracy of the ground state.

% \noindent

A chiral perturbation can be introduced by complementing (\ref{Hamiltonian}) with a term $-\delta\sum_i\left(\hat{\sigma}^{z}_i\hat{\tau}^{z}_{i+1}-\hat{\tau}^{z}_i\hat{\sigma}^{z}_{i+1}\right)$ \cite{schulz1983phase} to give
\begin{equation}\label{ChiralHamiltonian}
    \hat{H}=\hat{H}_0-\delta\sum_i\left(\hat{\sigma}^{z}_i\hat{\tau}^{z}_{i+1}-\hat{\tau}^{z}_i\hat{\sigma}^{z}_{i+1}\right).
\end{equation} 
At non-zero $\delta$ the model is no longer self-dual, and the location of the quantum phase transition as a function of $\beta$ is not known exactly. 
% Also, the introduced symmetries reduce to $\left[\hat{H}_0,\hat{P}_3 \right]=0$, breaking down the 4 fold ground state degeneracy of $\hat{H}_0$ into a 2 fold ground state degeneracy of $\hat{H}$. 
Regarding the nature of the transition, on theoretical grounds it is known that the crossover exponent $\phi$ for the chiral perturbation is given by \cite{schulz1983phase}
\begin{equation}
    \phi=\frac{3\nu}{2}+\frac{1}{4}-\frac{\nu^2}{2\nu-1}.
\end{equation}
Here, $\nu$ is the correlation length critical exponent which, in the case of vanishing chiral perturbation $\delta=0$, is known exactly as a function of the parameter $\lambda$,
\begin{equation}
    \nu=\frac{1}{2-\frac{\pi}{2}\left[\arccos\left(-\lambda\right)\right]^{-1}}.
\end{equation}
The chiral perturbation $\delta$ is relevant if $\phi>0$, which is the case for $\nu>\nu_c=\left(1+\sqrt{3}\right)/4\simeq0.683$.
This leads to a critical value of $\lambda$\cite{kohmoto,o2015symmetry},
\begin{equation}
\lambda_c=-\cos\frac{\pi\left(\sqrt{3}+1\right)}{4\left(\sqrt{3}-1\right)}\simeq0.9779,
\end{equation}
below which the chiral perturbation $\delta$ is always relevant. In the aforementioned numerical simulation of the Rydberg chain, it was found that away from the commensurate line, the melting of the period-four phase immediately occurs in the chiral universality class \cite{chepiga2021kibble} introduced by Huse and Fisher\cite{HuseFisher}. It is found that the conformal transition along the commensurate line is of the Ashkin-Teller type with a critical exponent of $\nu\simeq0.78$\cite{chepiga2021kibble,PhysRevResearch.4.043102}, i.e. with a value well into the regime where the chiral perturbation is relevant. Consistently it is found that away from the commensurate line, the chirality of the problem immediately drives the transition into the chiral universality class. Considering the Ashkin-Teller model however, there should be a region in the phase diagram ($\lambda_c<\lambda\leq1$) where $\delta$ is non-vanishing, but the C-IC transition remains in the Ashkin-Teller universality class. The phase diagram is presented in Figure \ref{fig:Diagram1}.

Originally, the Ashkin-Teller model was introduced as a classical model in statistical physics \cite{ashkin1943statistics}. Via the well established 2D classical to 1D quantum correspondence, the Hamiltonian of the quantum Ashkin-Teller model can be obtained by considering the highly anisotropic limit of the 2D classical model on a square lattice\cite{kohmoto} . 
 Numerical results on the classical ferromagnetic chiral Ashkin-Teller model on the square lattice obtained with a CTMRG algorithm\cite{nyckees2022commensurate} are in qualitative agreement with the phase diagram presented in Fig. \ref{fig:Diagram1}. In this paper we explore the vicinity of the four-state Potts point where the main field theory prediction that chiral perturbations are irrelevant is still waiting for numerical verification.

\subsubsection{Algorithm}

We address the problem numerically with state-of-the art density matrix renormalization group algorithm\cite{dmrg1,dmrg2,dmrg3,dmrg4}.
To obtain the ground state in MPS-form we use a two-site DMRG algorithm, see for instance Ref.\onlinecite{dmrg4}. The fact that we are dealing with a local Hilbert space of dimension $d=4$ significantly increases the complexity of the algorithm and in turn limits the maximum bond dimension $D$ that we can reach to $D_\mathrm{max}=200-500$. To ensure convergence both in bond dimension and in the number of sweeps, the bond dimension was increased after every half-sweep in steps of 20 to arrive at $D_\mathrm{max}$. We treat the wave-function as converged when the ground-state energy does not change more than by $10^{-6}$ during one sweep. 

A boundary term $-\epsilon\left(\hat{\sigma}_1^z+\hat{\sigma}_1^z\hat{\tau}_1^z+\hat{\sigma}_N^z+\hat{\sigma}_N^z\hat{\tau}_N^z\right)$, where $N$ is the system size and epsilon some positive constant of the order of the ground state energy density per site, was introduced to break the 4-fold degeneracy of the ground state and to make sure one ends up in a state with non-vanishing local magnetization $\left<\hat{\sigma_i^z}\right>$.

With this DMRG algorithm we could simulate quantum chains with sufficient accuracy to  distinguish chiral and conformal transitions in the vicinity of the four-state Potts point and in the presence of chiral perturbations.

\subsubsection{Distinguishing chiral transition and floating phase}

To distinguish the different types of phase transitions, one needs a quantity that reliably sets the transitions apart from one another. Huse and Fisher argue that such a quantity is provided by the product of the incommensurability wave vector $q$ and the correlation length $\xi$. Generally it is expected that the wave vector $q$ approaches the commensurate value $0$ with an exponent $\bar{\beta}$. While this exponent is not exactly known in the case of the Ashkin-Teller model, it is argued that $\bar{\beta}>\nu$ as soon as the transition is conformal. This implies that upon approaching the transition the product $q\times\xi$ will decay to 0. In contrast, for a chiral transition $\bar{\beta}=\nu$ and the product $q\times\xi$ remains finite upon approaching the transition. Since the chiral transition is direct, the exponent $\nu$ should be the same coming from either side of the transition. In the case of a floating phase, the system is critical, i.e. it has a diverging correlation length, while $q$ is still non-vanishing. This implies that on the incommensurate side one will observe $q\times\xi$ to be diverging towards the transition. Coming from the commensurate side, the transition will occur via a Pokrovsky-Talapov transition, implying a correlation length diverging with an exponent $\nu=\frac{1}{2}$ \cite{Pokrovsky_Talapov}.

\subsubsection{Extraction of the correlation length and of the wave vector}
Using the notation introduced in Eq. (\ref{Hamiltonian}), we consider the connected correlation function of the operators $\sigma^z_{i}$:
\begin{equation}
C^{\sigma}_{ij}=\left<\hat{\sigma}^z_i\hat{\sigma}^z_j\right>-\left<\hat{\sigma}^z_i\right>\left<\hat{\sigma}^z_j\right>,
\end{equation}
where $i$ and $j$ denote the different sites. The correlation length $\xi$ as well as the wave vector $q$ are extracted by fitting the correlation function to the Ornstein-Zenicke form\cite{ornstein_zernike},
\begin{equation}
 C^{\sigma}_{ij}\propto\frac{\exp\left(-\frac{|i-j|}{\xi}\right)}{\sqrt{|i-j|}}\cos\left(q|i-j|+\varphi\right).  
\end{equation}

\begin{figure}
\includegraphics[width=0.45\textwidth]{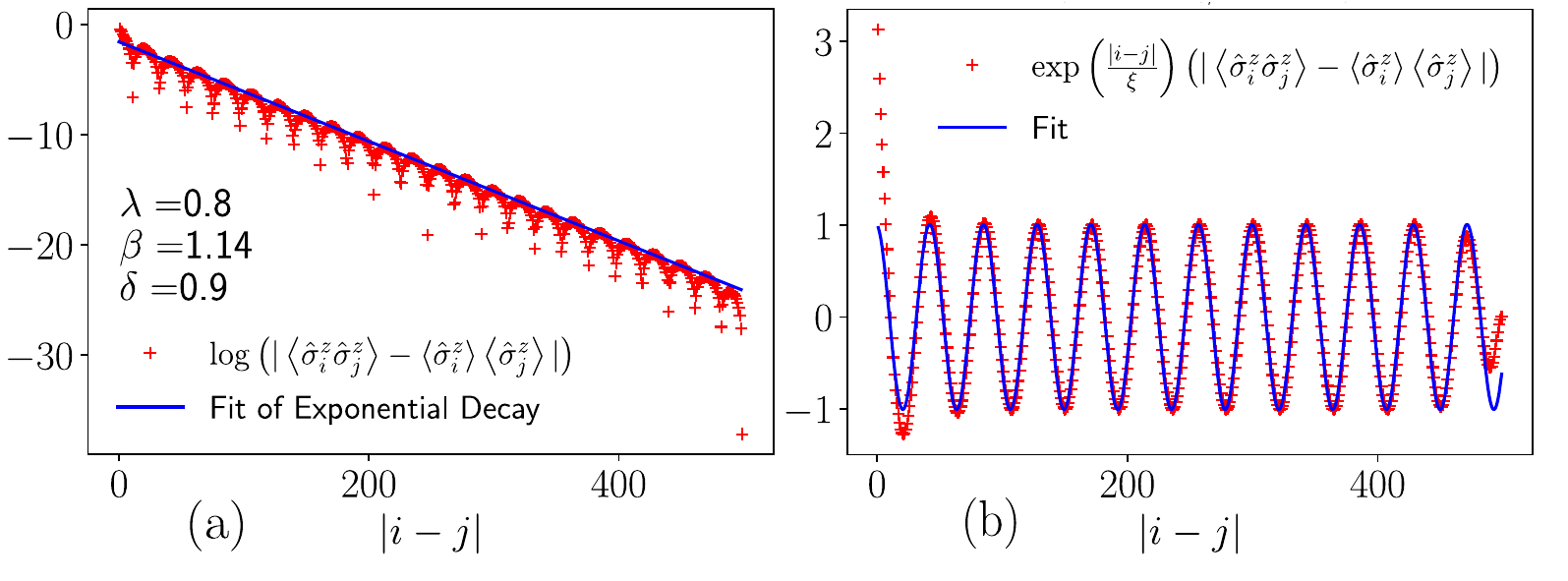}
\caption{Example of (a) the logarithm of $C^{\sigma}_{ij}$ as a function of the distance of two sites $|i-j|$ including a fit for the ``crest" of the oscillations and of (b) the corresponding rescaled correlation function, which is then used to extract $q$.}
\label{Xi_Rescaled_1000_08_114_09}
\end{figure}

This is attained by first fitting the correlation length, as is indicated in Figure (\ref{Xi_Rescaled_1000_08_114_09}a). When considering $\log\left(C^{\sigma}_{ij}\right)$, one easily sees that within the bulk, the maxima of the oscillations decay linearly as a function of $|i-j|$, with a slope of $-\frac{1}{\xi}$. 
Making use of this, one can then rescale $C^{\sigma}_{ij}$ with a factor $\exp\left(\frac{|i-j}{\xi}|\right)$, which then allows to fit the wave vector $q$, as indicated in Figure (\ref{Xi_Rescaled_1000_08_114_09}b).

\noindent

This procedure works very well for most of the points of the phase diagram. A problem arises when $q$ is very small, which is the case at the points close to the Ashkin-Teller line. The problem is due to the fact that for small $q$, the correlation function reaches machine precision before the elapse of a whole period, making it very difficult to properly fit the oscillations with a wave vector $q$. The way this issue was dealt with was by approaching the transition from a direction in parameter space where the oscillation period was smaller.

\subsubsection{Identifying the Lifshitz Line}
\begin{figure}
\centering
\includegraphics[width=0.5\textwidth]{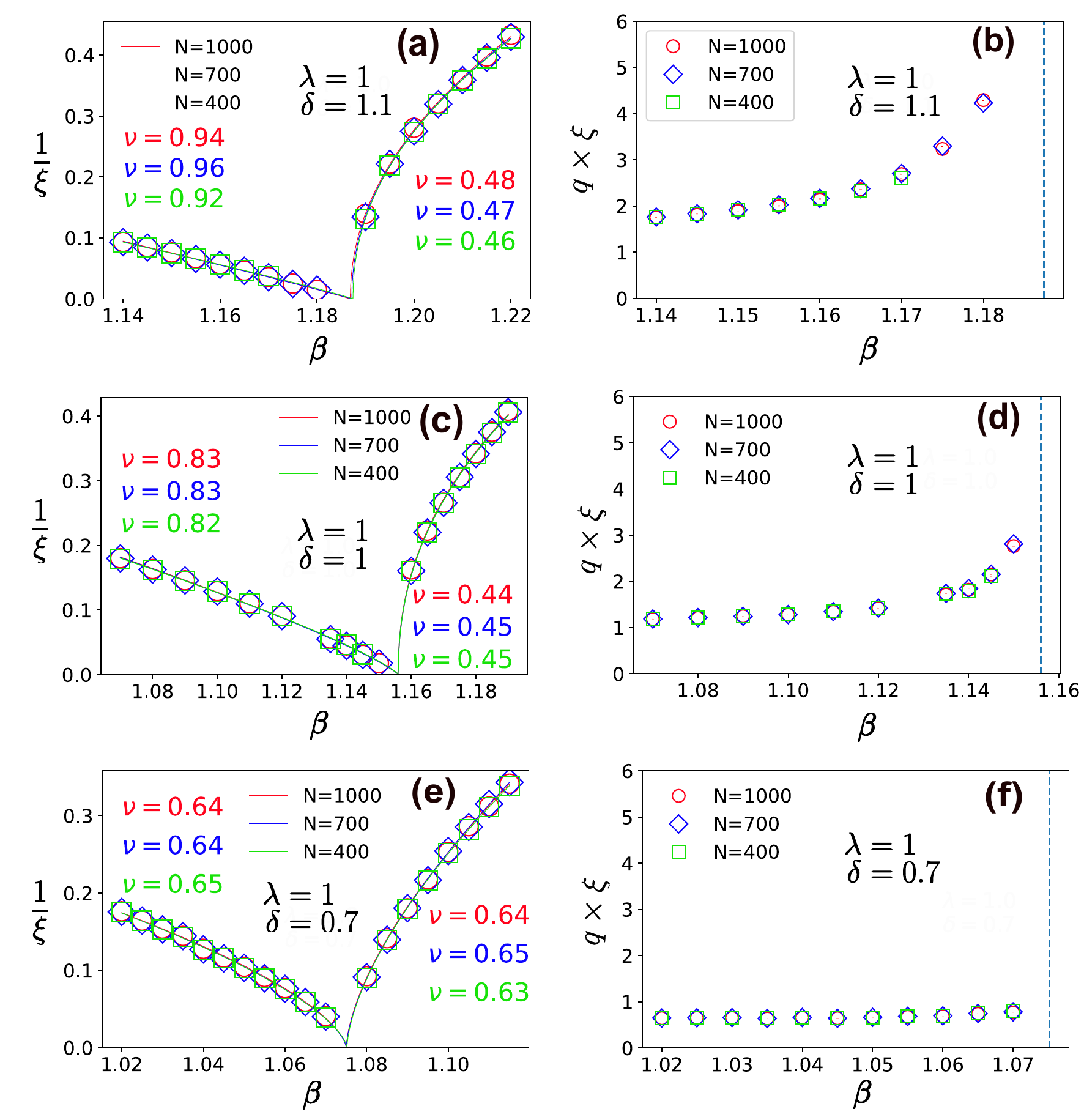}
\caption{Plots of $\frac{1}{\xi}$ and $q\times\xi$ around criticality at $\lambda=1$ and $\delta=1.1$ (a and b), as well as at $\delta=1.0$ (c and d) and $\delta=0.7$ (e and f). The simulations were carried out on three different system sizes $N=400$, $N=700$ and $N=1000$. In (a) and (b) one can see indications of an intermediate floating phase, namely the mismatch of the critical exponents $\nu_{left}$ and $\nu_{right}$ as well as a tendency to a convex behaviour close to the criticality coming from the left. There is also a clear increase in $q\times\xi$ close to the critical point. In (c) and (d) the results can be seen to be inconclusive as to whether the transition occurs in the chiral universality class or via a floating phase. In (e) and (f) everything points towards a chiral transition.}
\label{lambda1_Master}
\end{figure}

As already mentioned, the estimate of the Lifshitz line was obtained by scanning the phase diagram along the $\delta$-axis for given values of $\lambda$. The estimate was then made by considering the behaviour of $\xi$ and $q\times\xi$ at the transition for different values of $\delta$. We know that at a chiral transition, the critical exponents for the correlation length coming from either the commensurate or the incommensurate phase ($\nu_{left}$ and $\nu_{right}$) should match. Also if there is an intermediate floating phase there should be a change in curvature on the $\frac{1}{\xi}$-plot due to the scaling of the correlation length $\xi \sim\exp(\frac{c}{\sqrt{\beta-\beta_c}})$, as well a diverging behaviour in $q\times\xi$ close to criticality. Figure (\ref{lambda1_Master}) shows two plots along the $\delta$-axis for $\lambda=1$.
Whereas in the Figure (\ref{lambda1_Master}a) and (\ref{lambda1_Master}b) everything points toward a critical floating phase at $\lambda=1$, $\delta=1.1$, Figures (\ref{lambda1_Master}e) and (\ref{lambda1_Master}f) point towards a chiral transition at $\lambda=1$, $\delta=0.7$. Considerations like these enable one to map out the different phases and put an estimate, albeit a very approximate one, for the Lifshitz point.

%%%%%%%%%%%%%%%%%%%%%%%%%%%%%%%%%%%%%%%%%%%%%%%%%%%%%%%%%%%%%%%%%%%%%%%%%%%%%%%%%%%%%%%%%%%%%%%%%%%

\section{Results}
\label{sec:results}
\subsubsection{Phase Diagram}
\begin{figure}[h!]
\includegraphics[width=0.45\textwidth]{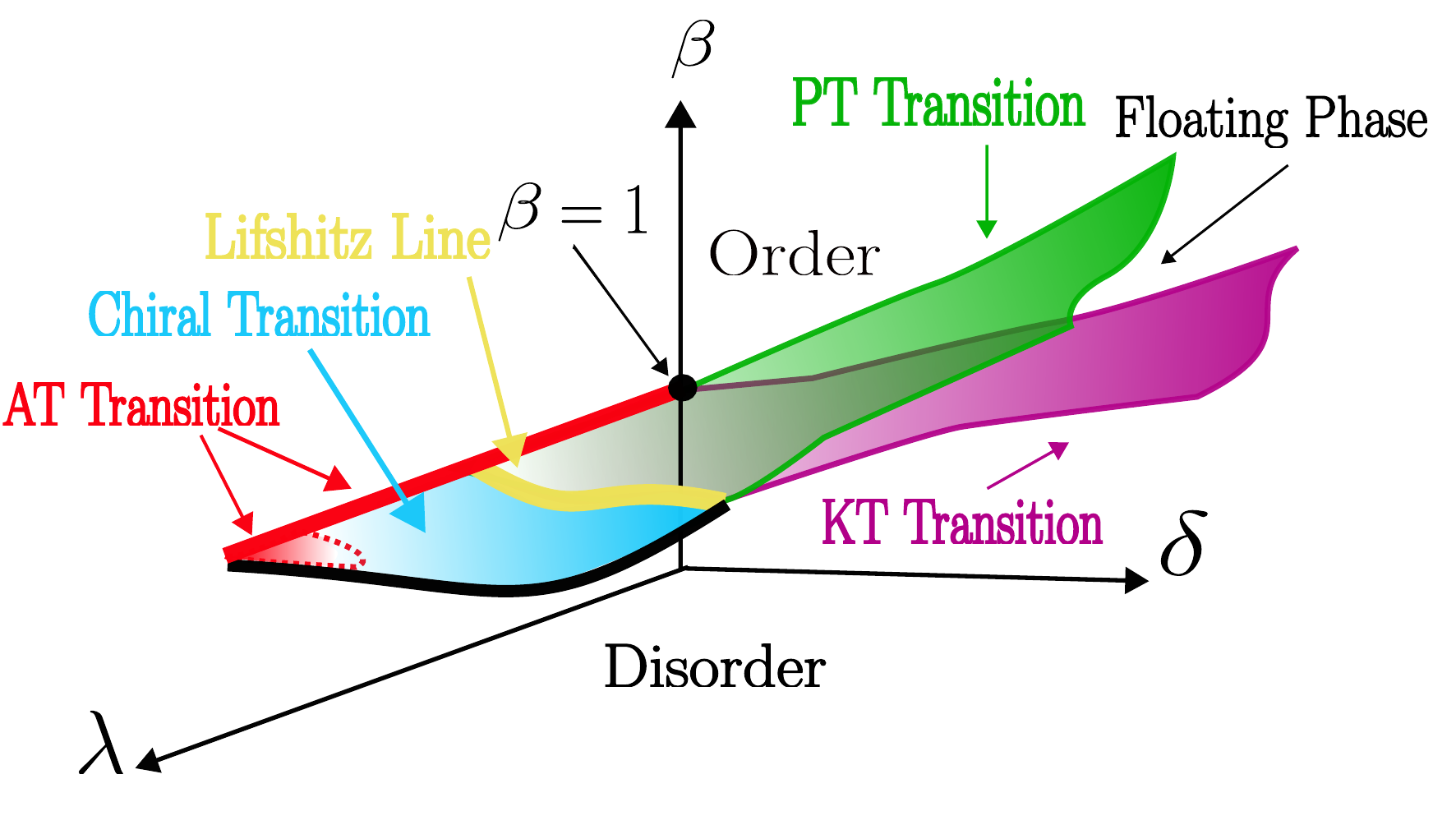}
\caption{Qualitative sketch of the nature of the quantum phase transitions between the ordered phase with a four-fold degenerate ground state and the disordered phase of the quantum Ashkin-Teller model defined by the Hamiltonian of Eq.(\ref{ChiralHamiltonian}) as a function of $\lambda$, $\beta$, and $\delta$.}
\label{PhaseDiagramSketch}
\end{figure}

The qualitative phase diagram we found is presented in Figure \ref{fig:Diagram1}. This diagram should be understood as follows. For every pair of parameters $(\lambda,\delta)$ displayed in the two-dimensional plot there is a particular value of the parameter $\beta$, introduced in (\ref{Hamiltonian}), where the C-IC transition occurs. One can understand Figure \ref{fig:Diagram1} as the depiction of the nature of phase transitions occurring on a two-dimensional critical surface embedded in a three-dimensional parameter space. This is further illustrated in Figure (\ref{PhaseDiagramSketch}).

Around $\lambda\simeq0.5$ we find that a non-vanishing value of $\delta$ immediately causes a floating phase to open up, whereas for $0.5\lesssim\lambda\lesssim0.98$ the transition remains direct but in the chiral universality class once a non-vanishing chiral perturbation is introduced. Furthermore, we identify a small region $\lambda_c<\lambda\leq 1$ in which the transition occurs in the Ashkin-Teller universality class for small but non-zero values of $\delta$.
\subsubsection{$\lambda=1$}
At $\lambda=1$ it is conjectured that with increasing $\delta$, different phase transitions can be observed. As mentioned above, for $0.9779\leq \lambda \leq 1$  it is expected that when introducing a small but finite perturbation $\delta$, the incommensurate-commensurate transition still occurs in the Ashkin-Teller universality class. To examine this, simulations were carried out at $\lambda=1$ for different values of $\delta$. For a given pair of $\lambda$ and $\delta$, the critical point $\beta_c$ was identified, and in its proximity the incommensurability wave vector $q$ as well as the correlation length $\xi$ were extracted in order to consider the quantity $q\times\xi$ close to the transition.
For $\lambda=1$ simulations across the critical surface were carried out for a multitude of values of $\delta$. Each critical point was approached at 3 different angles and the scaling of $q\times\xi$ of the different cuts was compared. The value of $q\times\xi$ for each cut at the transition ({\it intercept}) was extracted. The average intercept for the three cuts was then computed.

\begin{figure}[h!]
\includegraphics[width=0.5\textwidth]{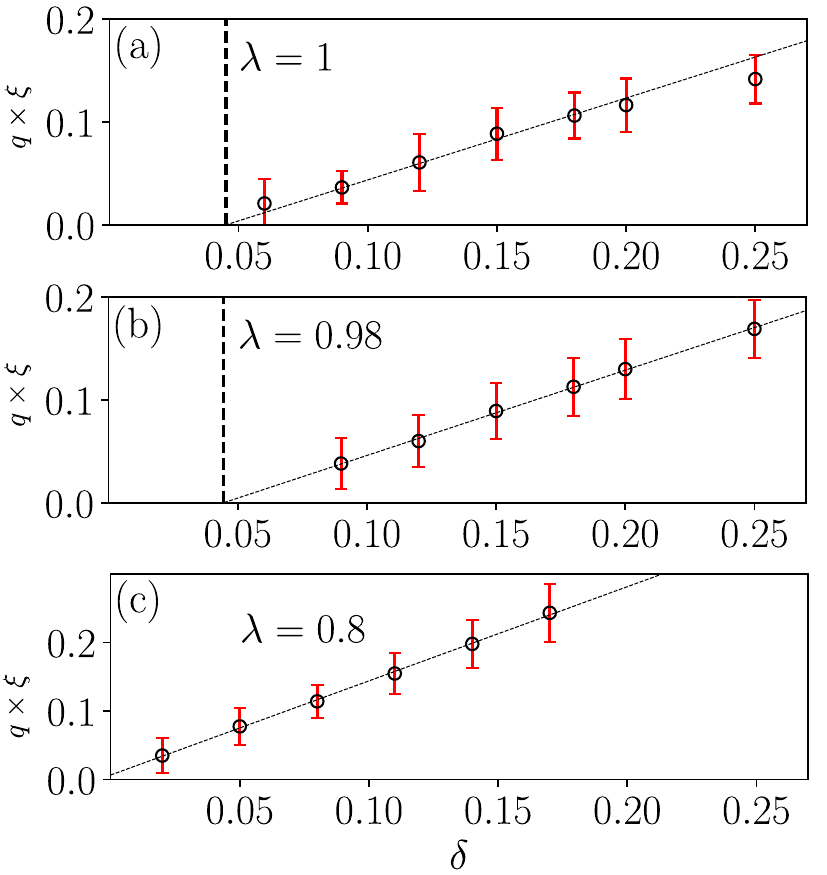}
\caption{$q\times\xi$ at the C-IC transition as a function of chiral coupling $\delta$ for (a) $\lambda=1$, (b) $\lambda=0.98$, and (c) $\lambda=0.8$. The dashed lines in (a) and (b) indicate the value of $\delta^c$ associated with the point where $q\times\xi$ vanishes and below which the transition is conformal. The dotted line is a linear fit.}
\label{Qxi_Intercepts_both_Lambda}
\end{figure}

\begin{figure}[h!]
\includegraphics[width=0.3\textwidth]{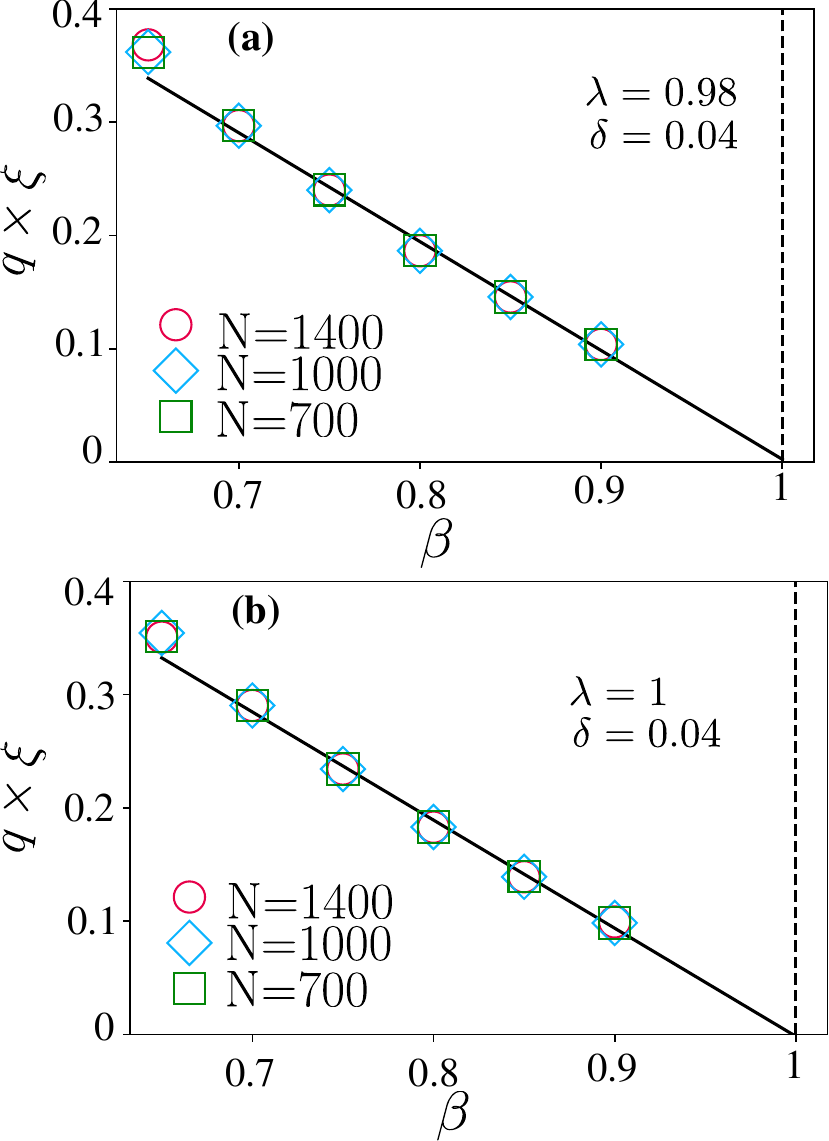}
\caption{The scaling of the quantity $q\times\xi$ with the distance to the C-IC transition at (a) $\lambda=0.98$, $\delta_{crit}=0.04$ and at (b) $\lambda=1$, $\delta_{crit}=0.04$  }
\label{Qxi_both_004_oblique3}
\end{figure}

As can be seen in Figure (\ref{Qxi_Intercepts_both_Lambda}a), as $\delta$ decreases, so does the value of the intercepts. Taking the errors into consideration, this suggests that $q\times\xi$ goes to zero around $\delta\simeq 0.05$. This would suggest that for $\delta$ smaller than this value, the incommensurate-commensurate transition actually occurs via an Ashkin-Teller type transition. In Figure (\ref{Qxi_both_004_oblique3}b) we can see $q\times\xi$ going to zero at the transition at $\lambda=1$ and $\delta=0.04$.

% \begin{figure}
% \centering
% \begin{subfigure}{0.23\textwidth}
% \includegraphics[width=\textwidth]{Plots/../Plots/Qxi_1.0_005_oblique3.eps}
% \caption{}
% \label{Qxi_1_005_oblique3}
% \end{subfigure}
% \hfill
% \begin{subfigure}{0.23\textwidth}
% \includegraphics[width=\textwidth]{Plots/../Plots/Qxi_0.98_005_oblique3.eps}
% \caption{}
% \label{Qxi_098_005_oblique3}
% \end{subfigure}
% \caption{The quantity $q\times\xi$ close to the transition at $\delta_{crit}=0.05$ for (a) $\lambda=1$ and (b) $\lambda=0.98$ for systems of size $N=1000$. The label "Oblique Cut" denotes the fact that the transition was not approached at a constant value of $\delta$, but rather the value of $\delta$ was uniformly decreased to reach $\delta=0.5$ at criticality.}
% \label{Qxi_both_Lambdas}
% \end{figure}

\subsubsection{$\lambda=0.98$}
The same procedure was carried out for $\lambda=0.98$. As for $\lambda=1$, we find $q\times\xi$ going to zero around $\delta\simeq 0.05$, Figure (\ref{Qxi_Intercepts_both_Lambda}b). Figure (\ref{Qxi_both_004_oblique3}a) seems to confirm that the incommensurate-commensurate transition for $\lambda=0.98$ and $\delta=0.04$ occurs in the Ashkin-Teller universality class. As a comparison, we carried out the same procedure for $\lambda=0.8$ where we know the chiral perturbation to be immediately relevant. In Figure (\ref{Qxi_Intercepts_both_Lambda}c) we see that, by contrast to $\lambda=1$ and $\lambda=0.98$, the quantity $q\times\xi$ at the transition only vanishes for $\delta=0$  

\subsubsection{Chiral Transition}
\begin{figure}
\includegraphics[width=0.45\textwidth]{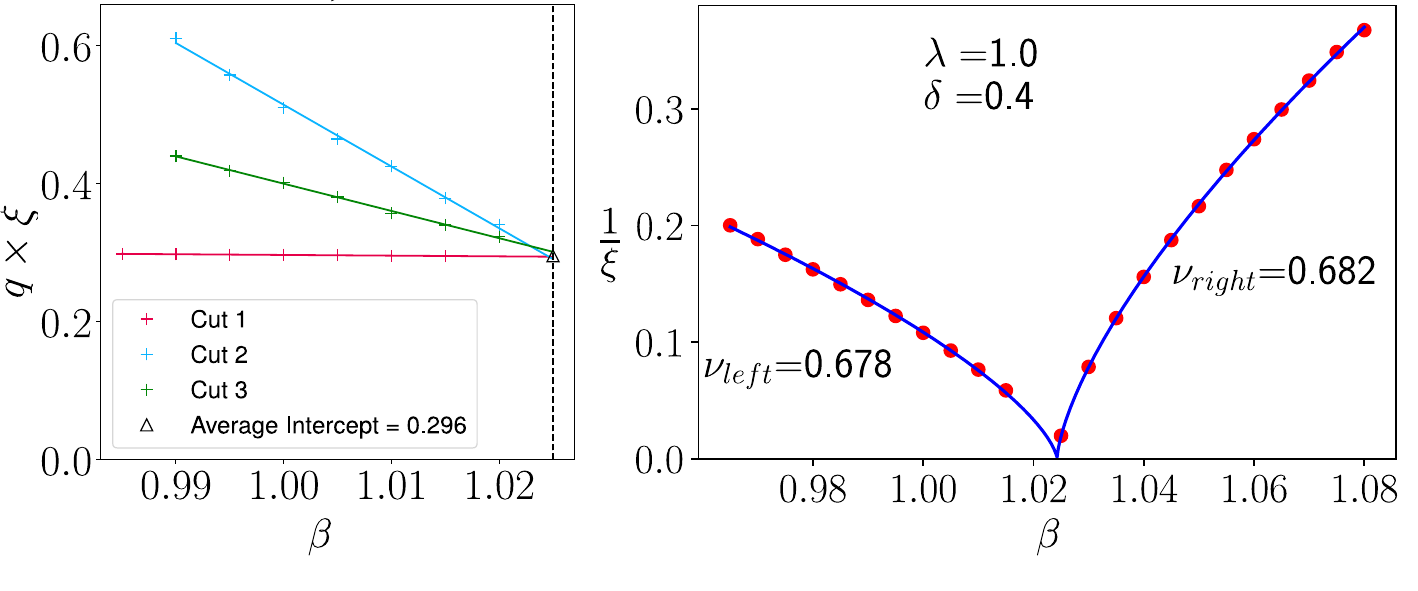}
\caption{(a) $q\times\xi$ as the IC-C transition is approached at $\lambda=1$, $\delta=0.4$. Every cut stands for a different direction in the plane spanned by the parameters ($\delta,\beta$) at which the transition is approached. (b) The inverse of the correlation length across the IC-C transition at $\lambda=1$, $\delta=0.4$. For the fit, the two exponents $\nu_{left}$ and $\nu_{right}$ as well as the point $\beta_{critical}$, at which the transition occurs, were fitted simultaneously.}
\label{Qxi_3cuts}
\end{figure}

It is conjectured that for $\lambda$ and $\delta$ large enough there will be a region in the phase diagram where the IC-C transition occurs via a chiral transition, characterized by the convergence of $q\times\xi$ upon approaching the transition. Indeed, the results of Section 1 seem to confirm this conjecture in the region of the phase diagram with $\lambda\simeq1$ and $\delta\geq0.05$. This can be examplified by considering the value $\delta=0.4$ at $\lambda=1$. This $\delta$-value is considerably larger than $\delta=0.05$, but still small enough that the transition does not occur via a floating phase.
The results of the simulation for $q\times\xi$ of three different cuts are shown in Figure (\ref{Qxi_3cuts}a). Independent of the angle at which the transition was approached in parameter space, $q\times\xi$ approaches the same value.

\noindent
To further substantiate the nature of this transition as chiral, the behaviour of the correlation length was considered on both sides of the transition. Figure (\ref{Qxi_3cuts}b) shows that the critical exponents $\nu$ of the correlation length for both sides of the transition match, as predicted by Huse and Fisher.

\subsubsection{Lifshitz Line}
For any value of $\lambda$ there is a $\delta$ large enough to open up a floating phase. Since every point where the floating phase opens up stands at the boundary of 3 different phases, we refer to the set of those points as Lifshitz line. An accurate location of the Lifshitz line is an extremely challenging task and the Lifshitz line presented in the Fig.\ref{fig:Diagram1} is defined up to an error $\Delta\delta \pm 0.05$.
In agreement with classical results, we observe that for $\lambda\lesssim 0.5$ a floating phase opens up immediately once a chiral perturbation $\delta$ is introduced. Also, the diagram indicates the small region close to $\lambda=1$ where we observe an Ashkin-Teller transition for finite $\delta$.
 
To obtain a qualitative idea of the phase diagram including the Lifshitz-line, simulations were also carried out for values of $\lambda$ smaller than $1$. The exact determination of the Lifshitz point for a given $\lambda$ is virtually impossible with the methods we used. One can however end up with a reasonable estimate by simply scanning the phase diagram. 

In Figure \ref{fig:Diagram1} one can see a qualitative diagram indicating the range in parameter space of the different types of transitions.

\section{Discussion}
\label{sec:discussion}
In the present paper we have investigated the critical properties of the quantum Ashkin-Teller model on a 1D chain with chiral perturbations. The resulting phase diagram is in qualitative agreement with the previously reported phase diagram of the classical symmetric 2D chiral Ashkin-Teller model on the square lattice. We find clear evidence of the presence of a region where the C-IC transition occurs in the chiral universality class predicted by Huse and Fisher. Consistent with the underlying theory we find that for $\lambda \lesssim 0.98$ a chiral perturbation immediately drives the C-IC transition out of the Ashkin-Teller universality class into either the chiral universality class ($0.5\lesssim\lambda$) or the melting via a critical floating phase ($\lambda\lesssim0.5$). We map out the location of the Lifshitz line indicating the boundary between the chiral transition and the floating phase. 
%The line qualitatively agrees with the results of the classical model, found with a Corner Transfer Matrix Renormalization Group algorithm.

Furthermore, our methods allowed us to examine the effect of a chiral perturbation in the region $0.98\lesssim\lambda\lesssim 1$, where, on theoretical grounds, it is known that the perturbation is irrelevant. Indeed we are able to identify a small domain in the phase diagram where we observe a C-IC transition in the Ashkin-Teller universality class for finite perturbations up to the value $\delta\simeq 0.04$. This region has so far not been observed in the 2D classical chiral Ashkin-Teller model. It would be interesting to see if the boundary of this phase indeed corresponds to the critical value $\nu=\nu_c=\left(1+\sqrt{3}\right)/4\simeq0.683$, but with our current algorithm the precision on $\nu$ is not sufficient to check this prediction, and this point is left for future investigation.

Finally, let us briefly comment on possible experimental realizations of this phase diagram with arrays of Rydberg atoms trapped in optical tweezers. As stated above, the conformal point at the boundary of the period-4 lobe in the simplest array of Rydberg atoms is characterized by the critical exponent $\nu\approx0.78$\cite{chepiga2021kibble,PhysRevResearch.4.043102}. However, recent proposal on multi-component and multi-species Rydberg arrays\cite{PhysRevX.12.011040,PhysRevLett.128.083202,Levi_2015,2023arXiv230812838C,2023npjQI...9...90C}, add new independent parameters that can be individually controlled in experiments. This opens a way to tune the conformal critical point and the Ashkin-Teller asymmetry parameter $\lambda$ and in turn to manipulate the appearance of the chiral transition at the boundary of the period-4 phase. A priori, it should be possible to realize the Ashkin-Teller critical point with $\lambda\gtrsim 0.98$ followed by a finite interval of conformal transitions. Kibble-Zurek experiments should in principle be able to check this scenario, but this will require to reach a much higher accuracy than that currently available.

\section*{Acknowledgments}
The authors acknowledge Samuel Nyckees for useful discussions. The work has been supported by the Swiss National Science Foundation (FM) grant 182179 and by the Delft Technology Fellowship (NC). Numerical simulations have been performed on the Dutch national e-infrastructure with the support of the SURF Cooperative, the facilities of the Scientific IT and Application Support Center of EPFL and at the DelftBlue HPC.

\appendix*
\section{Approaching a Transition from Different Angles in Parameter Space}

\begin{figure}
\includegraphics[width=0.45\textwidth]{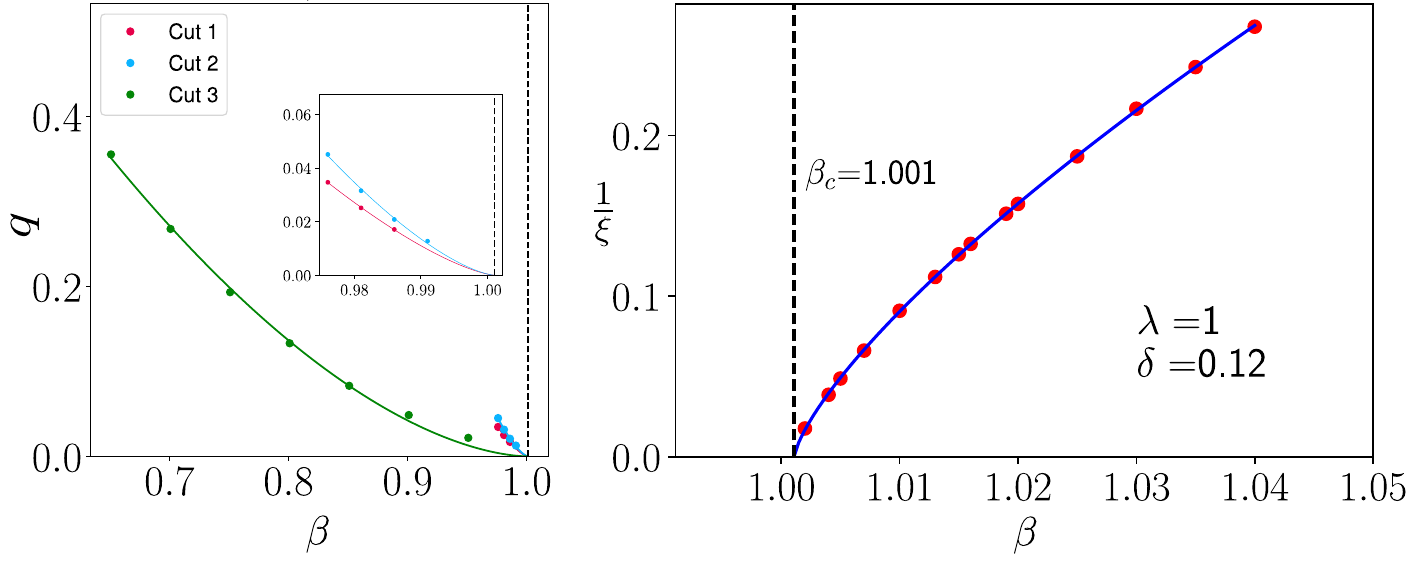}
\caption{(a) Determining $\beta_c$, i.e. the value in $\beta-$space where the C-IC transition occurs, by fitting the (inverse of the) correlation length with a power law. (b) The extracted wave vector $q$ for the three different cuts listed in Table \ref{DifferentCuts}. The inset shows an enlarged version of cuts $1$\&$2$ close to the transition.}
\label{Q_3cuts_1_012}
\end{figure}

The behaviour of the order parameters across the transition should not depend on the direction in parameter space along which the transition is approached. An illustrative example is given by the transition at $\lambda=1, \delta=0.12$. First, the point of the C-IC transition in $\beta$-space, labeled $\beta_c$, is determined by performing simulations at constant $\lambda$ and $\delta$ on the commensurate side and then fitting the diverging correlation length, Figure (\ref{Q_3cuts_1_012}b).

\noindent
Once $\beta_c$ has been determined, the transition is then approached from different angles in the $\beta-\delta$-plane. Table \ref{DifferentCuts} shows the three different cuts that were carried out in this particular example. The corresponding plots of $q$ are shown in Figure (\ref{Q_3cuts_1_012}a).
\begin{table}[!h]
\caption{Table of the points that were simulated to examine the behaviour around the transition at $\lambda=1$, $\delta=0.12$ and $\beta=1.001$. Each triplet in a row denotes a point that was simulated. As one can see from the bottom line in the table, each cut aims at the point of the transition, but it is approached from three different angles in the $\beta-\delta$-plane. $\lambda$ is always kept at $1$.}
\begin{tabular}{|lll|lll|lll|}
\hline
\multicolumn{3}{|l|}{Cut $1$}  & \multicolumn{3}{l|}{Cut $2$}   & \multicolumn{3}{l|}{Cut $3$}   \\ \hline
\hspace{0.2cm}$\lambda$\hspace{0.2cm} & \hspace{0.2cm}$\beta$ & \hspace{0.2cm}$\delta$ & \hspace{0.2cm}$\lambda$\hspace{0.2cm} & \hspace{0.2cm}$\beta$ & \hspace{0.2cm}$\delta$ & \hspace{0.2cm}$\lambda$\hspace{0.2cm} & \hspace{0.2cm}$\beta$ & \hspace{0.2cm}$\delta$ \\ \hline
& & & & & &\hspace{0.3cm}1         & 0.651   & 0.47 \\ 
 & & & & & &\hspace{0.3cm}1         & 0.701   & 0.42 \\ 
\hspace{0.3cm}1         & 0.976   & 0.32     & \hspace{0.3cm}1         & 0.976   & 0.37     & \hspace{0.3cm}1         & 0.751   & 0.37     \\
\hspace{0.3cm}1         & 0.981   & 0.28     & \hspace{0.3cm}1         & 0.981   & 0.32     & \hspace{0.3cm}1         & 0.801   & 0.32     \\
\hspace{0.3cm}1         & 0.986   & 0.24     & \hspace{0.3cm}1         & 0.986   & 0.27     & \hspace{0.3cm}1         & 0.851   & 0.27     \\
\hspace{0.3cm}1         & 0.991   & 0.2      & \hspace{0.3cm}1         & 0.991   & 0.22     & \hspace{0.3cm}1         & 0.901   & 0.22     \\
\hspace{0.3cm}1         & 0.996   & 0.16     & \hspace{0.3cm}1         & 0.996   & 0.17     & \hspace{0.3cm}1         & 0.951   & 0.17     \\
\hspace{0.3cm}1         & 1.001   & 0.12     & \hspace{0.3cm}1         & 1.001   & 0.12     & \hspace{0.3cm}1         & 1.001   & 0.12     \\ \hline
\end{tabular}
\label{DifferentCuts}
\end{table}

\bibliographystyle{apsrev4-1}
\bibliography{bibliography}

\begin{thebibliography}{39}%
\makeatletter
\providecommand \@ifxundefined [1]{%
 \@ifx{#1\undefined}
}%
\providecommand \@ifnum [1]{%
 \ifnum #1\expandafter \@firstoftwo
 \else \expandafter \@secondoftwo
 \fi
}%
\providecommand \@ifx [1]{%
 \ifx #1\expandafter \@firstoftwo
 \else \expandafter \@secondoftwo
 \fi
}%
\providecommand \natexlab [1]{#1}%
\providecommand \enquote  [1]{``#1''}%
\providecommand \bibnamefont  [1]{#1}%
\providecommand \bibfnamefont [1]{#1}%
\providecommand \citenamefont [1]{#1}%
\providecommand \href@noop [0]{\@secondoftwo}%
\providecommand \href [0]{\begingroup \@sanitize@url \@href}%
\providecommand \@href[1]{\@@startlink{#1}\@@href}%
\providecommand \@@href[1]{\endgroup#1\@@endlink}%
\providecommand \@sanitize@url [0]{\catcode `\\12\catcode `\$12\catcode
  `\&12\catcode `\#12\catcode `\^12\catcode `\_12\catcode `\%12\relax}%
\providecommand \@@startlink[1]{}%
\providecommand \@@endlink[0]{}%
\providecommand \url  [0]{\begingroup\@sanitize@url \@url }%
\providecommand \@url [1]{\endgroup\@href {#1}{\urlprefix }}%
\providecommand \urlprefix  [0]{URL }%
\providecommand \Eprint [0]{\href }%
\providecommand \doibase [0]{http://dx.doi.org/}%
\providecommand \selectlanguage [0]{\@gobble}%
\providecommand \bibinfo  [0]{\@secondoftwo}%
\providecommand \bibfield  [0]{\@secondoftwo}%
\providecommand \translation [1]{[#1]}%
\providecommand \BibitemOpen [0]{}%
\providecommand \bibitemStop [0]{}%
\providecommand \bibitemNoStop [0]{.\EOS\space}%
\providecommand \EOS [0]{\spacefactor3000\relax}%
\providecommand \BibitemShut  [1]{\csname bibitem#1\endcsname}%
\let\auto@bib@innerbib\@empty
%</preamble>
\bibitem [{\citenamefont {Giamarchi}(2004)}]{giamarchi}%
  \BibitemOpen
  \bibfield  {author} {\bibinfo {author} {\bibfnamefont {T.}~\bibnamefont
  {Giamarchi}},\ }\href {\doibase 10.1093/acprof:oso/9780198525004.001.0001}
  {\emph {\bibinfo {title} {{Quantum physics in one dimension}}}},\ Internat.
  Ser. Mono. Phys.\ (\bibinfo  {publisher} {Clarendon Press},\ \bibinfo
  {address} {Oxford},\ \bibinfo {year} {2004})\BibitemShut {NoStop}%
\bibitem [{\citenamefont {Tsvelik}(2003)}]{tsvelik}%
  \BibitemOpen
  \bibfield  {author} {\bibinfo {author} {\bibfnamefont {A.~M.}\ \bibnamefont
  {Tsvelik}},\ }\href {\doibase 10.1017/CBO9780511615832} {\emph {\bibinfo
  {title} {Quantum Field Theory in Condensed Matter Physics}}},\ \bibinfo
  {edition} {2nd}\ ed.\ (\bibinfo  {publisher} {Cambridge University Press},\
  \bibinfo {year} {2003})\BibitemShut {NoStop}%
\bibitem [{\citenamefont {Huse}\ and\ \citenamefont
  {Fisher}(1982)}]{HuseFisher}%
  \BibitemOpen
  \bibfield  {author} {\bibinfo {author} {\bibfnamefont {D.~A.}\ \bibnamefont
  {Huse}}\ and\ \bibinfo {author} {\bibfnamefont {M.~E.}\ \bibnamefont
  {Fisher}},\ }\href {\doibase 10.1103/PhysRevLett.49.793} {\bibfield
  {journal} {\bibinfo  {journal} {Phys. Rev. Lett.}\ }\textbf {\bibinfo
  {volume} {49}},\ \bibinfo {pages} {793} (\bibinfo {year} {1982})}\BibitemShut
  {NoStop}%
\bibitem [{\citenamefont {Selke}\ and\ \citenamefont
  {Yeomans}(1982)}]{Selke1982}%
  \BibitemOpen
  \bibfield  {author} {\bibinfo {author} {\bibfnamefont {W.}~\bibnamefont
  {Selke}}\ and\ \bibinfo {author} {\bibfnamefont {J.~M.}\ \bibnamefont
  {Yeomans}},\ }\href {\doibase 10.1007/BF01307706} {\bibfield  {journal}
  {\bibinfo  {journal} {Zeitschrift f{\"u}r Physik B Condensed Matter}\
  }\textbf {\bibinfo {volume} {46}},\ \bibinfo {pages} {311} (\bibinfo {year}
  {1982})}\BibitemShut {NoStop}%
\bibitem [{\citenamefont {Schulz}(1983)}]{schulz1983phase}%
  \BibitemOpen
  \bibfield  {author} {\bibinfo {author} {\bibfnamefont {H.}~\bibnamefont
  {Schulz}},\ }\href@noop {} {\bibfield  {journal} {\bibinfo  {journal}
  {Physical Review B}\ }\textbf {\bibinfo {volume} {28}},\ \bibinfo {pages}
  {2746} (\bibinfo {year} {1983})}\BibitemShut {NoStop}%
\bibitem [{\citenamefont {Haldane}\ \emph {et~al.}(1983)\citenamefont
  {Haldane}, \citenamefont {Bak},\ and\ \citenamefont {Bohr}}]{haldane_bak}%
  \BibitemOpen
  \bibfield  {author} {\bibinfo {author} {\bibfnamefont {F.~D.~M.}\
  \bibnamefont {Haldane}}, \bibinfo {author} {\bibfnamefont {P.}~\bibnamefont
  {Bak}}, \ and\ \bibinfo {author} {\bibfnamefont {T.}~\bibnamefont {Bohr}},\
  }\href {\doibase 10.1103/PhysRevB.28.2743} {\bibfield  {journal} {\bibinfo
  {journal} {Phys. Rev. B}\ }\textbf {\bibinfo {volume} {28}},\ \bibinfo
  {pages} {2743} (\bibinfo {year} {1983})}\BibitemShut {NoStop}%
\bibitem [{\citenamefont {Huse}\ and\ \citenamefont
  {Fisher}(1984)}]{HuseFisher1984}%
  \BibitemOpen
  \bibfield  {author} {\bibinfo {author} {\bibfnamefont {D.~A.}\ \bibnamefont
  {Huse}}\ and\ \bibinfo {author} {\bibfnamefont {M.~E.}\ \bibnamefont
  {Fisher}},\ }\href {\doibase 10.1103/PhysRevB.29.239} {\bibfield  {journal}
  {\bibinfo  {journal} {Phys. Rev. B}\ }\textbf {\bibinfo {volume} {29}},\
  \bibinfo {pages} {239} (\bibinfo {year} {1984})}\BibitemShut {NoStop}%
\bibitem [{\citenamefont {Schreiner}\ \emph {et~al.}(1994)\citenamefont
  {Schreiner}, \citenamefont {Jacobi},\ and\ \citenamefont
  {Selke}}]{SelkeExperiment}%
  \BibitemOpen
  \bibfield  {author} {\bibinfo {author} {\bibfnamefont {J.}~\bibnamefont
  {Schreiner}}, \bibinfo {author} {\bibfnamefont {K.}~\bibnamefont {Jacobi}}, \
  and\ \bibinfo {author} {\bibfnamefont {W.}~\bibnamefont {Selke}},\ }\href
  {\doibase 10.1103/PhysRevB.49.2706} {\bibfield  {journal} {\bibinfo
  {journal} {Phys. Rev. B}\ }\textbf {\bibinfo {volume} {49}},\ \bibinfo
  {pages} {2706} (\bibinfo {year} {1994})}\BibitemShut {NoStop}%
\bibitem [{\citenamefont {Pokrovsky}\ and\ \citenamefont
  {Talapov}(1979)}]{Pokrovsky_Talapov}%
  \BibitemOpen
  \bibfield  {author} {\bibinfo {author} {\bibfnamefont {V.~L.}\ \bibnamefont
  {Pokrovsky}}\ and\ \bibinfo {author} {\bibfnamefont {A.~L.}\ \bibnamefont
  {Talapov}},\ }\href {\doibase 10.1103/PhysRevLett.42.65} {\bibfield
  {journal} {\bibinfo  {journal} {Phys. Rev. Lett.}\ }\textbf {\bibinfo
  {volume} {42}},\ \bibinfo {pages} {65} (\bibinfo {year} {1979})}\BibitemShut
  {NoStop}%
\bibitem [{\citenamefont {Kosterlitz}\ and\ \citenamefont
  {Thouless}(1973)}]{Kosterlitz_Thouless}%
  \BibitemOpen
  \bibfield  {author} {\bibinfo {author} {\bibfnamefont {J.~M.}\ \bibnamefont
  {Kosterlitz}}\ and\ \bibinfo {author} {\bibfnamefont {D.~J.}\ \bibnamefont
  {Thouless}},\ }\href {http://stacks.iop.org/0022-3719/6/i=7/a=010} {\bibfield
   {journal} {\bibinfo  {journal} {Journal of Physics C: Solid State Physics}\
  }\textbf {\bibinfo {volume} {6}},\ \bibinfo {pages} {1181} (\bibinfo {year}
  {1973})}\BibitemShut {NoStop}%
\bibitem [{\citenamefont {Di~Francesco}\ \emph {et~al.}(1997)\citenamefont
  {Di~Francesco}, \citenamefont {Mathieu},\ and\ \citenamefont
  {S{\'e}n{\'e}chal}}]{difrancesco}%
  \BibitemOpen
  \bibfield  {author} {\bibinfo {author} {\bibfnamefont {P.}~\bibnamefont
  {Di~Francesco}}, \bibinfo {author} {\bibfnamefont {P.}~\bibnamefont
  {Mathieu}}, \ and\ \bibinfo {author} {\bibfnamefont {D.}~\bibnamefont
  {S{\'e}n{\'e}chal}},\ }\href {https://books.google.ch/books?id=keUrdME5rhIC}
  {\emph {\bibinfo {title} {Conformal Field Theory}}},\ Graduate Texts in
  Contemporary Physics\ (\bibinfo  {publisher} {Springer},\ \bibinfo {address}
  {New York},\ \bibinfo {year} {1997})\BibitemShut {NoStop}%
\bibitem [{\citenamefont {den Nijs}(1988)}]{Den_Nijs}%
  \BibitemOpen
  \bibfield  {author} {\bibinfo {author} {\bibfnamefont {M.}~\bibnamefont {den
  Nijs}},\ }\href@noop {} {\bibfield  {journal} {\bibinfo  {journal} {Phase
  Transitions and Critical Phenomena}\ }\textbf {\bibinfo {volume} {12}},\
  \bibinfo {pages} {219} (\bibinfo {year} {1988})}\BibitemShut {NoStop}%
\bibitem [{\citenamefont {Ostlund}(1981)}]{Ostlund}%
  \BibitemOpen
  \bibfield  {author} {\bibinfo {author} {\bibfnamefont {S.}~\bibnamefont
  {Ostlund}},\ }\href {\doibase 10.1103/PhysRevB.24.398} {\bibfield  {journal}
  {\bibinfo  {journal} {Phys. Rev. B}\ }\textbf {\bibinfo {volume} {24}},\
  \bibinfo {pages} {398} (\bibinfo {year} {1981})}\BibitemShut {NoStop}%
\bibitem [{\citenamefont {Nelson}\ and\ \citenamefont
  {Halperin}(1979)}]{nelson1979dislocation}%
  \BibitemOpen
  \bibfield  {author} {\bibinfo {author} {\bibfnamefont {D.~R.}\ \bibnamefont
  {Nelson}}\ and\ \bibinfo {author} {\bibfnamefont {B.}~\bibnamefont
  {Halperin}},\ }\href@noop {} {\bibfield  {journal} {\bibinfo  {journal}
  {Physical Review B}\ }\textbf {\bibinfo {volume} {19}},\ \bibinfo {pages}
  {2457} (\bibinfo {year} {1979})}\BibitemShut {NoStop}%
\bibitem [{\citenamefont {Chepiga}\ and\ \citenamefont
  {Mila}(2021{\natexlab{a}})}]{chepiga2021kibble}%
  \BibitemOpen
  \bibfield  {author} {\bibinfo {author} {\bibfnamefont {N.}~\bibnamefont
  {Chepiga}}\ and\ \bibinfo {author} {\bibfnamefont {F.}~\bibnamefont {Mila}},\
  }\href@noop {} {\bibfield  {journal} {\bibinfo  {journal} {Nature
  Communications}\ }\textbf {\bibinfo {volume} {12}},\ \bibinfo {pages} {1}
  (\bibinfo {year} {2021}{\natexlab{a}})}\BibitemShut {NoStop}%
\bibitem [{\citenamefont {Nyckees}\ and\ \citenamefont
  {Mila}(2022)}]{nyckees2022commensurate}%
  \BibitemOpen
  \bibfield  {author} {\bibinfo {author} {\bibfnamefont {S.}~\bibnamefont
  {Nyckees}}\ and\ \bibinfo {author} {\bibfnamefont {F.}~\bibnamefont {Mila}},\
  }\href@noop {} {\bibfield  {journal} {\bibinfo  {journal} {Physical Review
  Research}\ }\textbf {\bibinfo {volume} {4}},\ \bibinfo {pages} {013093}
  (\bibinfo {year} {2022})}\BibitemShut {NoStop}%
\bibitem [{\citenamefont {Bernien}\ \emph {et~al.}(2017)\citenamefont
  {Bernien}, \citenamefont {Schwartz}, \citenamefont {Keesling}, \citenamefont
  {Levine}, \citenamefont {Omran}, \citenamefont {Pichler}, \citenamefont
  {Choi}, \citenamefont {Zibrov}, \citenamefont {Endres}, \citenamefont
  {Greiner} \emph {et~al.}}]{bernien2017probing}%
  \BibitemOpen
  \bibfield  {author} {\bibinfo {author} {\bibfnamefont {H.}~\bibnamefont
  {Bernien}}, \bibinfo {author} {\bibfnamefont {S.}~\bibnamefont {Schwartz}},
  \bibinfo {author} {\bibfnamefont {A.}~\bibnamefont {Keesling}}, \bibinfo
  {author} {\bibfnamefont {H.}~\bibnamefont {Levine}}, \bibinfo {author}
  {\bibfnamefont {A.}~\bibnamefont {Omran}}, \bibinfo {author} {\bibfnamefont
  {H.}~\bibnamefont {Pichler}}, \bibinfo {author} {\bibfnamefont
  {S.}~\bibnamefont {Choi}}, \bibinfo {author} {\bibfnamefont {A.~S.}\
  \bibnamefont {Zibrov}}, \bibinfo {author} {\bibfnamefont {M.}~\bibnamefont
  {Endres}}, \bibinfo {author} {\bibfnamefont {M.}~\bibnamefont {Greiner}},
  \emph {et~al.},\ }\href@noop {} {\bibfield  {journal} {\bibinfo  {journal}
  {Nature}\ }\textbf {\bibinfo {volume} {551}},\ \bibinfo {pages} {579}
  (\bibinfo {year} {2017})}\BibitemShut {NoStop}%
\bibitem [{\citenamefont {Keesling}\ \emph {et~al.}(2019)\citenamefont
  {Keesling}, \citenamefont {Omran}, \citenamefont {Levine}, \citenamefont
  {Bernien}, \citenamefont {Pichler}, \citenamefont {Choi}, \citenamefont
  {Samajdar}, \citenamefont {Schwartz}, \citenamefont {Silvi}, \citenamefont
  {Sachdev} \emph {et~al.}}]{keesling2019quantum}%
  \BibitemOpen
  \bibfield  {author} {\bibinfo {author} {\bibfnamefont {A.}~\bibnamefont
  {Keesling}}, \bibinfo {author} {\bibfnamefont {A.}~\bibnamefont {Omran}},
  \bibinfo {author} {\bibfnamefont {H.}~\bibnamefont {Levine}}, \bibinfo
  {author} {\bibfnamefont {H.}~\bibnamefont {Bernien}}, \bibinfo {author}
  {\bibfnamefont {H.}~\bibnamefont {Pichler}}, \bibinfo {author} {\bibfnamefont
  {S.}~\bibnamefont {Choi}}, \bibinfo {author} {\bibfnamefont {R.}~\bibnamefont
  {Samajdar}}, \bibinfo {author} {\bibfnamefont {S.}~\bibnamefont {Schwartz}},
  \bibinfo {author} {\bibfnamefont {P.}~\bibnamefont {Silvi}}, \bibinfo
  {author} {\bibfnamefont {S.}~\bibnamefont {Sachdev}},  \emph {et~al.},\
  }\href@noop {} {\bibfield  {journal} {\bibinfo  {journal} {Nature}\ }\textbf
  {\bibinfo {volume} {568}},\ \bibinfo {pages} {207} (\bibinfo {year}
  {2019})}\BibitemShut {NoStop}%
\bibitem [{\citenamefont {Rader}\ and\ \citenamefont
  {L\"auchli}(2019)}]{rader2019floating}%
  \BibitemOpen
  \bibfield  {author} {\bibinfo {author} {\bibfnamefont {M.}~\bibnamefont
  {Rader}}\ and\ \bibinfo {author} {\bibfnamefont {A.~M.}\ \bibnamefont
  {L\"auchli}},\ }\href@noop {} {\enquote {\bibinfo {title} {Floating phases in
  one-dimensional rydberg ising chains},}\ } (\bibinfo {year} {2019}),\ \Eprint
  {http://arxiv.org/abs/1908.02068} {arXiv:1908.02068 [cond-mat.quant-gas]}
  \BibitemShut {NoStop}%
\bibitem [{\citenamefont {Whitsitt}\ \emph
  {et~al.}(2018{\natexlab{a}})\citenamefont {Whitsitt}, \citenamefont
  {Samajdar},\ and\ \citenamefont {Sachdev}}]{PhysRevB.98.205118}%
  \BibitemOpen
  \bibfield  {author} {\bibinfo {author} {\bibfnamefont {S.}~\bibnamefont
  {Whitsitt}}, \bibinfo {author} {\bibfnamefont {R.}~\bibnamefont {Samajdar}},
  \ and\ \bibinfo {author} {\bibfnamefont {S.}~\bibnamefont {Sachdev}},\ }\href
  {\doibase 10.1103/PhysRevB.98.205118} {\bibfield  {journal} {\bibinfo
  {journal} {Phys. Rev. B}\ }\textbf {\bibinfo {volume} {98}},\ \bibinfo
  {pages} {205118} (\bibinfo {year} {2018}{\natexlab{a}})}\BibitemShut
  {NoStop}%
\bibitem [{\citenamefont {Chepiga}\ and\ \citenamefont
  {Mila}(2019{\natexlab{a}})}]{prl_chepiga}%
  \BibitemOpen
  \bibfield  {author} {\bibinfo {author} {\bibfnamefont {N.}~\bibnamefont
  {Chepiga}}\ and\ \bibinfo {author} {\bibfnamefont {F.}~\bibnamefont {Mila}},\
  }\href {\doibase 10.1103/PhysRevLett.122.017205} {\bibfield  {journal}
  {\bibinfo  {journal} {Phys. Rev. Lett.}\ }\textbf {\bibinfo {volume} {122}},\
  \bibinfo {pages} {017205} (\bibinfo {year} {2019}{\natexlab{a}})}\BibitemShut
  {NoStop}%
\bibitem [{\citenamefont {Chepiga}\ and\ \citenamefont
  {Mila}(2019{\natexlab{b}})}]{scipost_chepiga}%
  \BibitemOpen
  \bibfield  {author} {\bibinfo {author} {\bibfnamefont {N.}~\bibnamefont
  {Chepiga}}\ and\ \bibinfo {author} {\bibfnamefont {F.}~\bibnamefont {Mila}},\
  }\href {\doibase 10.21468/SciPostPhys.6.3.033} {\bibfield  {journal}
  {\bibinfo  {journal} {SciPost Phys.}\ }\textbf {\bibinfo {volume} {6}},\
  \bibinfo {pages} {33} (\bibinfo {year} {2019}{\natexlab{b}})}\BibitemShut
  {NoStop}%
\bibitem [{\citenamefont {Giudici}\ \emph {et~al.}(2019)\citenamefont
  {Giudici}, \citenamefont {Angelone}, \citenamefont {Magnifico}, \citenamefont
  {Zeng}, \citenamefont {Giudice}, \citenamefont {Mendes-Santos},\ and\
  \citenamefont {Dalmonte}}]{dalmonte}%
  \BibitemOpen
  \bibfield  {author} {\bibinfo {author} {\bibfnamefont {G.}~\bibnamefont
  {Giudici}}, \bibinfo {author} {\bibfnamefont {A.}~\bibnamefont {Angelone}},
  \bibinfo {author} {\bibfnamefont {G.}~\bibnamefont {Magnifico}}, \bibinfo
  {author} {\bibfnamefont {Z.}~\bibnamefont {Zeng}}, \bibinfo {author}
  {\bibfnamefont {G.}~\bibnamefont {Giudice}}, \bibinfo {author} {\bibfnamefont
  {T.}~\bibnamefont {Mendes-Santos}}, \ and\ \bibinfo {author} {\bibfnamefont
  {M.}~\bibnamefont {Dalmonte}},\ }\href {\doibase 10.1103/PhysRevB.99.094434}
  {\bibfield  {journal} {\bibinfo  {journal} {Phys. Rev. B}\ }\textbf {\bibinfo
  {volume} {99}},\ \bibinfo {pages} {094434} (\bibinfo {year}
  {2019})}\BibitemShut {NoStop}%
\bibitem [{\citenamefont {Whitsitt}\ \emph
  {et~al.}(2018{\natexlab{b}})\citenamefont {Whitsitt}, \citenamefont
  {Samajdar},\ and\ \citenamefont {Sachdev}}]{sachdev_dual}%
  \BibitemOpen
  \bibfield  {author} {\bibinfo {author} {\bibfnamefont {S.}~\bibnamefont
  {Whitsitt}}, \bibinfo {author} {\bibfnamefont {R.}~\bibnamefont {Samajdar}},
  \ and\ \bibinfo {author} {\bibfnamefont {S.}~\bibnamefont {Sachdev}},\ }\href
  {\doibase 10.1103/PhysRevB.98.205118} {\bibfield  {journal} {\bibinfo
  {journal} {Phys. Rev. B}\ }\textbf {\bibinfo {volume} {98}},\ \bibinfo
  {pages} {205118} (\bibinfo {year} {2018}{\natexlab{b}})}\BibitemShut
  {NoStop}%
\bibitem [{\citenamefont {Chepiga}\ and\ \citenamefont
  {Mila}(2021{\natexlab{b}})}]{PhysRevResearch.3.023049}%
  \BibitemOpen
  \bibfield  {author} {\bibinfo {author} {\bibfnamefont {N.}~\bibnamefont
  {Chepiga}}\ and\ \bibinfo {author} {\bibfnamefont {F.}~\bibnamefont {Mila}},\
  }\href {\doibase 10.1103/PhysRevResearch.3.023049} {\bibfield  {journal}
  {\bibinfo  {journal} {Phys. Rev. Res.}\ }\textbf {\bibinfo {volume} {3}},\
  \bibinfo {pages} {023049} (\bibinfo {year} {2021}{\natexlab{b}})}\BibitemShut
  {NoStop}%
\bibitem [{\citenamefont {Maceira}\ \emph {et~al.}(2022)\citenamefont
  {Maceira}, \citenamefont {Chepiga},\ and\ \citenamefont
  {Mila}}]{PhysRevResearch.4.043102}%
  \BibitemOpen
  \bibfield  {author} {\bibinfo {author} {\bibfnamefont {I.~A.}\ \bibnamefont
  {Maceira}}, \bibinfo {author} {\bibfnamefont {N.}~\bibnamefont {Chepiga}}, \
  and\ \bibinfo {author} {\bibfnamefont {F.}~\bibnamefont {Mila}},\ }\href
  {\doibase 10.1103/PhysRevResearch.4.043102} {\bibfield  {journal} {\bibinfo
  {journal} {Phys. Rev. Res.}\ }\textbf {\bibinfo {volume} {4}},\ \bibinfo
  {pages} {043102} (\bibinfo {year} {2022})}\BibitemShut {NoStop}%
\bibitem [{\citenamefont {Kohmoto}\ \emph {et~al.}(1981)\citenamefont
  {Kohmoto}, \citenamefont {den Nijs},\ and\ \citenamefont
  {Kadanoff}}]{kohmoto}%
  \BibitemOpen
  \bibfield  {author} {\bibinfo {author} {\bibfnamefont {M.}~\bibnamefont
  {Kohmoto}}, \bibinfo {author} {\bibfnamefont {M.}~\bibnamefont {den Nijs}}, \
  and\ \bibinfo {author} {\bibfnamefont {L.~P.}\ \bibnamefont {Kadanoff}},\
  }\href {\doibase 10.1103/PhysRevB.24.5229} {\bibfield  {journal} {\bibinfo
  {journal} {Phys. Rev. B}\ }\textbf {\bibinfo {volume} {24}},\ \bibinfo
  {pages} {5229} (\bibinfo {year} {1981})}\BibitemShut {NoStop}%
\bibitem [{\citenamefont {O'Brien}\ \emph {et~al.}(2015)\citenamefont
  {O'Brien}, \citenamefont {Bartlett}, \citenamefont {Doherty},\ and\
  \citenamefont {Flammia}}]{o2015symmetry}%
  \BibitemOpen
  \bibfield  {author} {\bibinfo {author} {\bibfnamefont {A.}~\bibnamefont
  {O'Brien}}, \bibinfo {author} {\bibfnamefont {S.~D.}\ \bibnamefont
  {Bartlett}}, \bibinfo {author} {\bibfnamefont {A.~C.}\ \bibnamefont
  {Doherty}}, \ and\ \bibinfo {author} {\bibfnamefont {S.~T.}\ \bibnamefont
  {Flammia}},\ }\href@noop {} {\bibfield  {journal} {\bibinfo  {journal}
  {Physical Review E}\ }\textbf {\bibinfo {volume} {92}},\ \bibinfo {pages}
  {042163} (\bibinfo {year} {2015})}\BibitemShut {NoStop}%
\bibitem [{\citenamefont {Ashkin}\ and\ \citenamefont
  {Teller}(1943)}]{ashkin1943statistics}%
  \BibitemOpen
  \bibfield  {author} {\bibinfo {author} {\bibfnamefont {J.}~\bibnamefont
  {Ashkin}}\ and\ \bibinfo {author} {\bibfnamefont {E.}~\bibnamefont
  {Teller}},\ }\href@noop {} {\bibfield  {journal} {\bibinfo  {journal}
  {Physical Review}\ }\textbf {\bibinfo {volume} {64}},\ \bibinfo {pages} {178}
  (\bibinfo {year} {1943})}\BibitemShut {NoStop}%
\bibitem [{\citenamefont {White}(1992)}]{dmrg1}%
  \BibitemOpen
  \bibfield  {author} {\bibinfo {author} {\bibfnamefont {S.~R.}\ \bibnamefont
  {White}},\ }\href {\doibase 10.1103/PhysRevLett.69.2863} {\bibfield
  {journal} {\bibinfo  {journal} {Phys. Rev. Lett.}\ }\textbf {\bibinfo
  {volume} {69}},\ \bibinfo {pages} {2863} (\bibinfo {year}
  {1992})}\BibitemShut {NoStop}%
\bibitem [{\citenamefont {Schollw\"ock}(2005)}]{dmrg2}%
  \BibitemOpen
  \bibfield  {author} {\bibinfo {author} {\bibfnamefont {U.}~\bibnamefont
  {Schollw\"ock}},\ }\href {\doibase 10.1103/RevModPhys.77.259} {\bibfield
  {journal} {\bibinfo  {journal} {Rev. Mod. Phys.}\ }\textbf {\bibinfo {volume}
  {77}},\ \bibinfo {pages} {259} (\bibinfo {year} {2005})}\BibitemShut
  {NoStop}%
\bibitem [{\citenamefont {\"Ostlund}\ and\ \citenamefont
  {Rommer}(1995)}]{dmrg3}%
  \BibitemOpen
  \bibfield  {author} {\bibinfo {author} {\bibfnamefont {S.}~\bibnamefont
  {\"Ostlund}}\ and\ \bibinfo {author} {\bibfnamefont {S.}~\bibnamefont
  {Rommer}},\ }\href {\doibase 10.1103/PhysRevLett.75.3537} {\bibfield
  {journal} {\bibinfo  {journal} {Phys. Rev. Lett.}\ }\textbf {\bibinfo
  {volume} {75}},\ \bibinfo {pages} {3537} (\bibinfo {year}
  {1995})}\BibitemShut {NoStop}%
\bibitem [{\citenamefont {Schollw\"ock}(2011)}]{dmrg4}%
  \BibitemOpen
  \bibfield  {author} {\bibinfo {author} {\bibfnamefont {U.}~\bibnamefont
  {Schollw\"ock}},\ }\href {\doibase
  http://dx.doi.org/10.1016/j.aop.2010.09.012} {\bibfield  {journal} {\bibinfo
  {journal} {Annals of Physics}\ }\textbf {\bibinfo {volume} {326}},\ \bibinfo
  {pages} {96 } (\bibinfo {year} {2011})}\BibitemShut {NoStop}%
\bibitem [{\citenamefont {Ornstein}\ and\ \citenamefont
  {Zernike}(1914)}]{ornstein_zernike}%
  \BibitemOpen
  \bibfield  {author} {\bibinfo {author} {\bibfnamefont {L.}~\bibnamefont
  {Ornstein}}\ and\ \bibinfo {author} {\bibfnamefont {F.}~\bibnamefont
  {Zernike}},\ }\href@noop {} {\bibfield  {journal} {\bibinfo  {journal}
  {Koninklijke Nederlandse Akademie van Wetenschappen Proceedings Series B
  Physical Sciences}\ }\textbf {\bibinfo {volume} {17}},\ \bibinfo {pages}
  {793} (\bibinfo {year} {1914})}\BibitemShut {NoStop}%
\bibitem [{\citenamefont {Singh}\ \emph {et~al.}(2022)\citenamefont {Singh},
  \citenamefont {Anand}, \citenamefont {Pocklington}, \citenamefont {Kemp},\
  and\ \citenamefont {Bernien}}]{PhysRevX.12.011040}%
  \BibitemOpen
  \bibfield  {author} {\bibinfo {author} {\bibfnamefont {K.}~\bibnamefont
  {Singh}}, \bibinfo {author} {\bibfnamefont {S.}~\bibnamefont {Anand}},
  \bibinfo {author} {\bibfnamefont {A.}~\bibnamefont {Pocklington}}, \bibinfo
  {author} {\bibfnamefont {J.~T.}\ \bibnamefont {Kemp}}, \ and\ \bibinfo
  {author} {\bibfnamefont {H.}~\bibnamefont {Bernien}},\ }\href {\doibase
  10.1103/PhysRevX.12.011040} {\bibfield  {journal} {\bibinfo  {journal} {Phys.
  Rev. X}\ }\textbf {\bibinfo {volume} {12}},\ \bibinfo {pages} {011040}
  (\bibinfo {year} {2022})}\BibitemShut {NoStop}%
\bibitem [{\citenamefont {Sheng}\ \emph {et~al.}(2022)\citenamefont {Sheng},
  \citenamefont {Hou}, \citenamefont {He}, \citenamefont {Wang}, \citenamefont
  {Guo}, \citenamefont {Zhuang}, \citenamefont {Mamat}, \citenamefont {Xu},
  \citenamefont {Liu}, \citenamefont {Wang},\ and\ \citenamefont
  {Zhan}}]{PhysRevLett.128.083202}%
  \BibitemOpen
  \bibfield  {author} {\bibinfo {author} {\bibfnamefont {C.}~\bibnamefont
  {Sheng}}, \bibinfo {author} {\bibfnamefont {J.}~\bibnamefont {Hou}}, \bibinfo
  {author} {\bibfnamefont {X.}~\bibnamefont {He}}, \bibinfo {author}
  {\bibfnamefont {K.}~\bibnamefont {Wang}}, \bibinfo {author} {\bibfnamefont
  {R.}~\bibnamefont {Guo}}, \bibinfo {author} {\bibfnamefont {J.}~\bibnamefont
  {Zhuang}}, \bibinfo {author} {\bibfnamefont {B.}~\bibnamefont {Mamat}},
  \bibinfo {author} {\bibfnamefont {P.}~\bibnamefont {Xu}}, \bibinfo {author}
  {\bibfnamefont {M.}~\bibnamefont {Liu}}, \bibinfo {author} {\bibfnamefont
  {J.}~\bibnamefont {Wang}}, \ and\ \bibinfo {author} {\bibfnamefont
  {M.}~\bibnamefont {Zhan}},\ }\href {\doibase 10.1103/PhysRevLett.128.083202}
  {\bibfield  {journal} {\bibinfo  {journal} {Phys. Rev. Lett.}\ }\textbf
  {\bibinfo {volume} {128}},\ \bibinfo {pages} {083202} (\bibinfo {year}
  {2022})}\BibitemShut {NoStop}%
\bibitem [{\citenamefont {Levi}\ \emph {et~al.}(2015)\citenamefont {Levi},
  \citenamefont {Minář}, \citenamefont {Garrahan},\ and\ \citenamefont
  {Lesanovsky}}]{Levi_2015}%
  \BibitemOpen
  \bibfield  {author} {\bibinfo {author} {\bibfnamefont {E.}~\bibnamefont
  {Levi}}, \bibinfo {author} {\bibfnamefont {J.}~\bibnamefont {Minář}},
  \bibinfo {author} {\bibfnamefont {J.~P.}\ \bibnamefont {Garrahan}}, \ and\
  \bibinfo {author} {\bibfnamefont {I.}~\bibnamefont {Lesanovsky}},\ }\href
  {\doibase 10.1088/1367-2630/17/12/123017} {\bibfield  {journal} {\bibinfo
  {journal} {New Journal of Physics}\ }\textbf {\bibinfo {volume} {17}},\
  \bibinfo {pages} {123017} (\bibinfo {year} {2015})}\BibitemShut {NoStop}%
\bibitem [{\citenamefont {{Chepiga}}(2023)}]{2023arXiv230812838C}%
  \BibitemOpen
  \bibfield  {author} {\bibinfo {author} {\bibfnamefont {N.}~\bibnamefont
  {{Chepiga}}},\ }\href {\doibase 10.48550/arXiv.2308.12838} {\bibfield
  {journal} {\bibinfo  {journal} {arXiv e-prints}\ ,\ \bibinfo {eid}
  {arXiv:2308.12838}} (\bibinfo {year} {2023})},\ \Eprint
  {http://arxiv.org/abs/2308.12838} {arXiv:2308.12838} \BibitemShut {NoStop}%
\bibitem [{\citenamefont {{Covey}}\ \emph {et~al.}(2023)\citenamefont
  {{Covey}}, \citenamefont {{Weinfurter}},\ and\ \citenamefont
  {{Bernien}}}]{2023npjQI...9...90C}%
  \BibitemOpen
  \bibfield  {author} {\bibinfo {author} {\bibfnamefont {J.~P.}\ \bibnamefont
  {{Covey}}}, \bibinfo {author} {\bibfnamefont {H.}~\bibnamefont
  {{Weinfurter}}}, \ and\ \bibinfo {author} {\bibfnamefont {H.}~\bibnamefont
  {{Bernien}}},\ }\href {\doibase 10.1038/s41534-023-00759-9} {\bibfield
  {journal} {\bibinfo  {journal} {npj Quantum Information}\ }\textbf {\bibinfo
  {volume} {9}},\ \bibinfo {eid} {90} (\bibinfo {year} {2023})}\BibitemShut
  {NoStop}%
\end{thebibliography}%

\end{document}